\documentclass[twocolumn,twoside,slac_two]{revtex4}
\usepackage{graphicx}
\usepackage{fancyhdr}
\usepackage{graphics}
\usepackage{epstopdf}
\usepackage{textpos}
\newcommand{\dedxp}{\ensuremath{\mathrm{d}E\!/\!\mathrm{d}x_{\rm Pixel}}}
\newcommand{\betat}{\ensuremath{ {\beta}_{\mathrm{Tile}}}}

\begin{document}

%%%%%%%%%%%%%%%%%%%%%% WRITE THE TITLE HERE %%%%%%%%%%%%%%%%%%%
\title{\centering Searches for Long-lived particles with the ATLAS experiment}
%%%%%%%%%%%%%%%%%%%%%% WRITE THE AUTHOR HERE %%%%%%%%%%%%%%%%%

%%% Please insert your personal picture here!

\author{
\centering
\begin{center}
Paul D. Jackson, on behalf of the ATLAS collaboration
\end{center}}
\affiliation{\centering SLAC National Accelerator Laboratory, 2575 Sand Hill Road, Menlo Park, California 94025, USA}
%%%%%%%%%%%%%%%%%%%%%% WRITE THE ABSTRACT HERE %%%%%%%%%%%%%%%%
\begin{abstract}
The discovery of a new type of a heavy long-lived particle (LLP) would be of fundamental significance due to their existence in many beyond the Standard Model scenarios~\cite{bib:fairbarn}.
LLPs are anticipated in a wide range of physics models which extend the Standard Model, such as Supersymmetry and Universal Extra Dimensions. 
Since LLPs produced in 7 TeV $pp$ collisions at the CERN LHC can be slow ($\beta << $1) and penetrating, time-of-flight and anomalous d$E$/dx energy 
loss measurements are promising ways to search for LLPs. In some cases these heavy objects may lose all of their energy and come to rest 
within the densest parts of the detector volume, decaying later, potentially out-of-time with collisions from the LHC. We present 
searches for LLPs using the ATLAS experiment, describing the techniques used and the results achieved to date.
\end{abstract}

\maketitle
\thispagestyle{fancy}

\section{\label{sec:introduction}Introduction}

The commencement of proton-proton physics at the Large Hadron Collider (LHC) in 2010
opened a new window on exploration at the high energy frontier. Sensitivity to new physics processes
are pushed beyond that of previous machines by studying the first few inverse picobarns of data
collected in $pp$ collisions at $\sqrt{s}$~=7~TeV.
Long-lived particles (LLPs)~\cite{bib:fairbarn} provide unique detector signatures which are free
of Standard Model physics backgrounds. Herein, the results 
of early searches~\cite{bib:llp1,bib:llp2} performed with the ATLAS 
experiment~\cite{bib:atlas} at the CERN LHC
utilizing data samples of 34-37~pb$^{-1}$ collected in 2010 are summarized.

\section{\label{sec:analysis}Analysis Strategies}

\noindent LLPs exhibiting lepton-like properties (such as long-lived sleptons) are expected to lose energy through electromagnetic processes. 
One would expect their detector signatures to resemble that of slow-moving muons. Hadron-like LLPs, such as R-hadrons, 
could exchange electric charge while penetrating the detector. It is possible, therefore, that they may be dominantly neutral in the 
inner tracker or the muon spectrometer~\cite{bib:farrar}. Two search strategies were pursued. 

\subsection{\label{subsec:muon_agnostic}Inner Detector/Calorimeter Based Search}

This search~\cite{bib:llp1} is designed to focus on hadron-like LLPs and combines information from the inner detector (ID) and calorimetery of ATLAS.
Events are triggered and selected with $E^{\rm{miss}}_{T}~>~$40~GeV. Candidates are required to be associated with high momentum tracks
($p_{T}~>~$50~GeV) and be central in the detector ($|\eta|~<~$1.7). Two independent technniques are then used to estimated the speed
of the candidate particle. The candidate $\beta\gamma$ is extracted from measurement of the ionization energy loss in the pixel detector,
the innermost tracking detector of ATLAS. Time-of-flight measurements are performed using the tile calorimeter, for energy deposits which 
match an ID track.
Track candidates are extrapolated to the tile calorimeter and the timing information of the matching cells is used in providing an additional 
estimate of the $\beta$ of the candidate. 
A combination of the track momentum with the two independent speed estimations yields two mass estimates using the relation $m~=~p/\beta\gamma$.   
The resulting two-dimensional plane defines the signal search regions. In order to estimate the yield of background processes contributing
to this region data-driven methods are used to extrapolate from the low mass regions into the higher mass regions where we would be sensitive to
any potential signals.

\begin{figure}[tbp]
%\ \!\!\!\!\!\!\!\!
\centering
\includegraphics[width=0.5\textwidth]{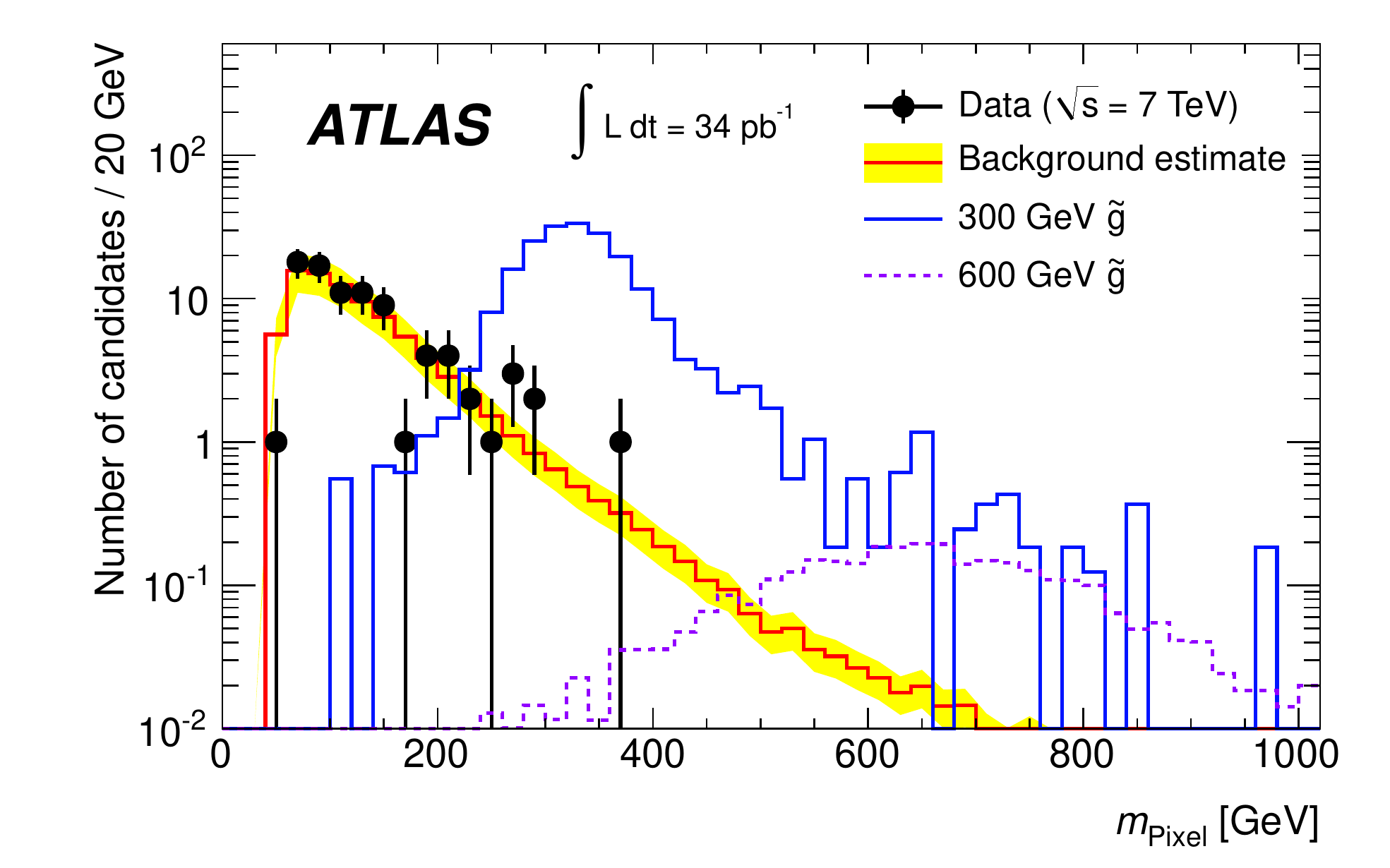}\\%\!\!\!\!\!\!
\includegraphics[width=0.5\textwidth]{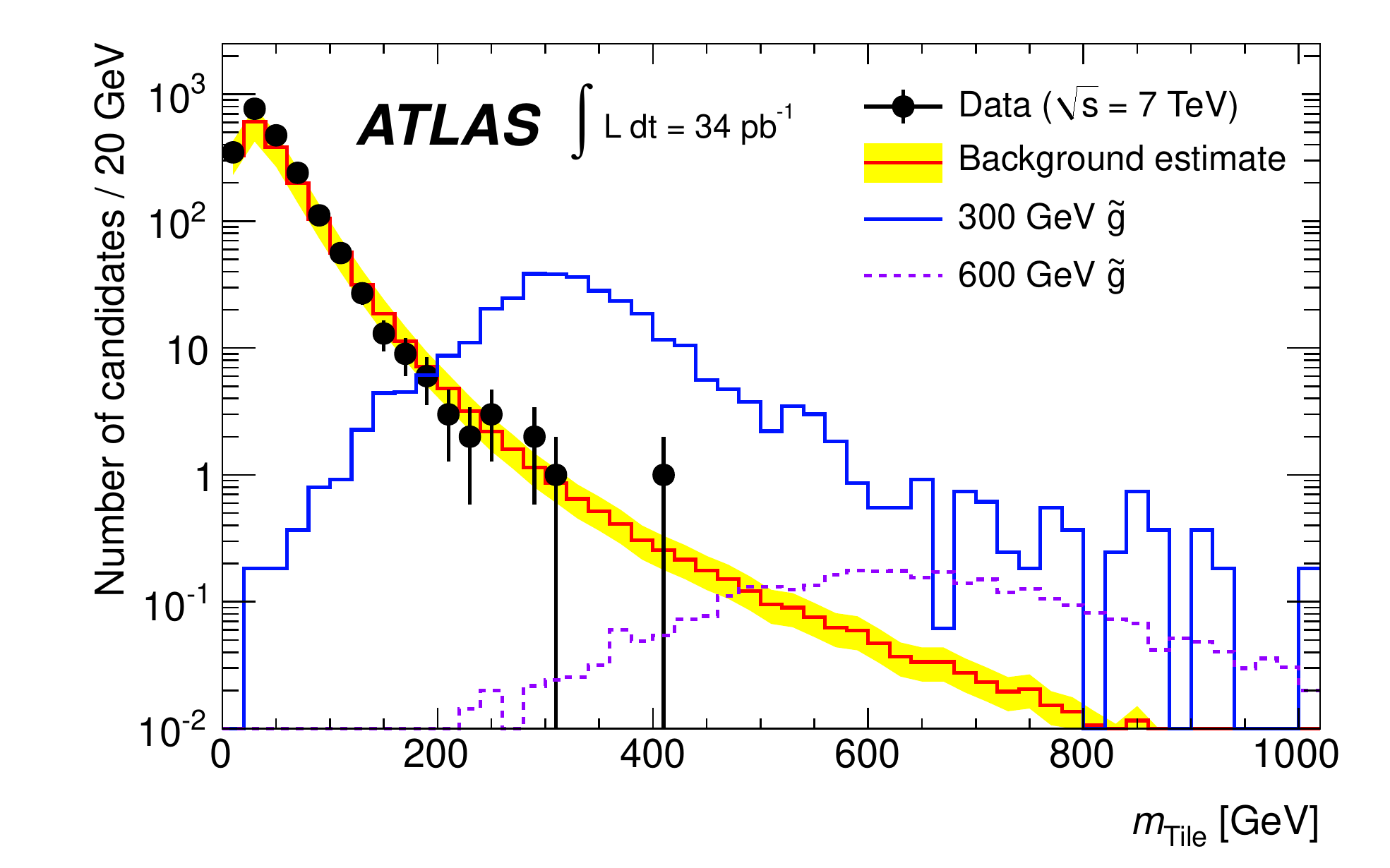}
\caption{Background estimates for the pixel detector (upper) and the tile calorimeter (lower)~\cite{bib:llp1}.
Signal samples are superimposed on the background estimate. The total systematic uncertainty of the background estimate is indicated by the error band.}
\label{fig:bkg-est}
\end{figure}

No significant correlations between the measurements of momentum, \dedxp, and \betat{} are observed. This is exploited to estimate the amount of 
background arising from instrumental effects.
Estimates for the background distributions of the mass estimates are obtained by combining random momentum values (after the kinematic 
cuts defined above) with random measurements of \dedxp{} and \betat.
The sampling is performed from candidates passing the kinematic selection criteria for the case of \betat{}, while \dedxp{} is extracted 
from a sample fulfilling $10<p_T<20$~GeV.
The process is repeated many times to reduce fluctuations and the resulting estimates are normalised to match the number of 
events in data. The resulting background estimates can be seen in Figure~\ref{fig:bkg-est} for the pixel detector 
(requiring $\dedxp > 1.8$~GeV$\rm{g}^{-1}\rm{cm}^{2}$) and the tile calorimeter (requiring $\betat<1$) 
separately. As can be seen from the Figures, there is a good overall agreement between the distribution of candidates in data and 
the background estimate. The expected background at high mass is generally small.

Combining the pixel detector and the tile calorimeter mass estimates further reduces
the background while retaining most of the expected signal (for more detail see~\cite{bib:llp1}).
In contrast to the individual background estimates shown in 
Figure \ref{fig:bkg-est}, the combined background is obtained by combining one random momentum value with 
random measurements of both \dedxp{} and $\betat$.
\subsection{\label{subsec:muon}Muon Spectrometer Based Search}
\begin{figure}[tbp]
  \centering
  \includegraphics[width=0.86\columnwidth]{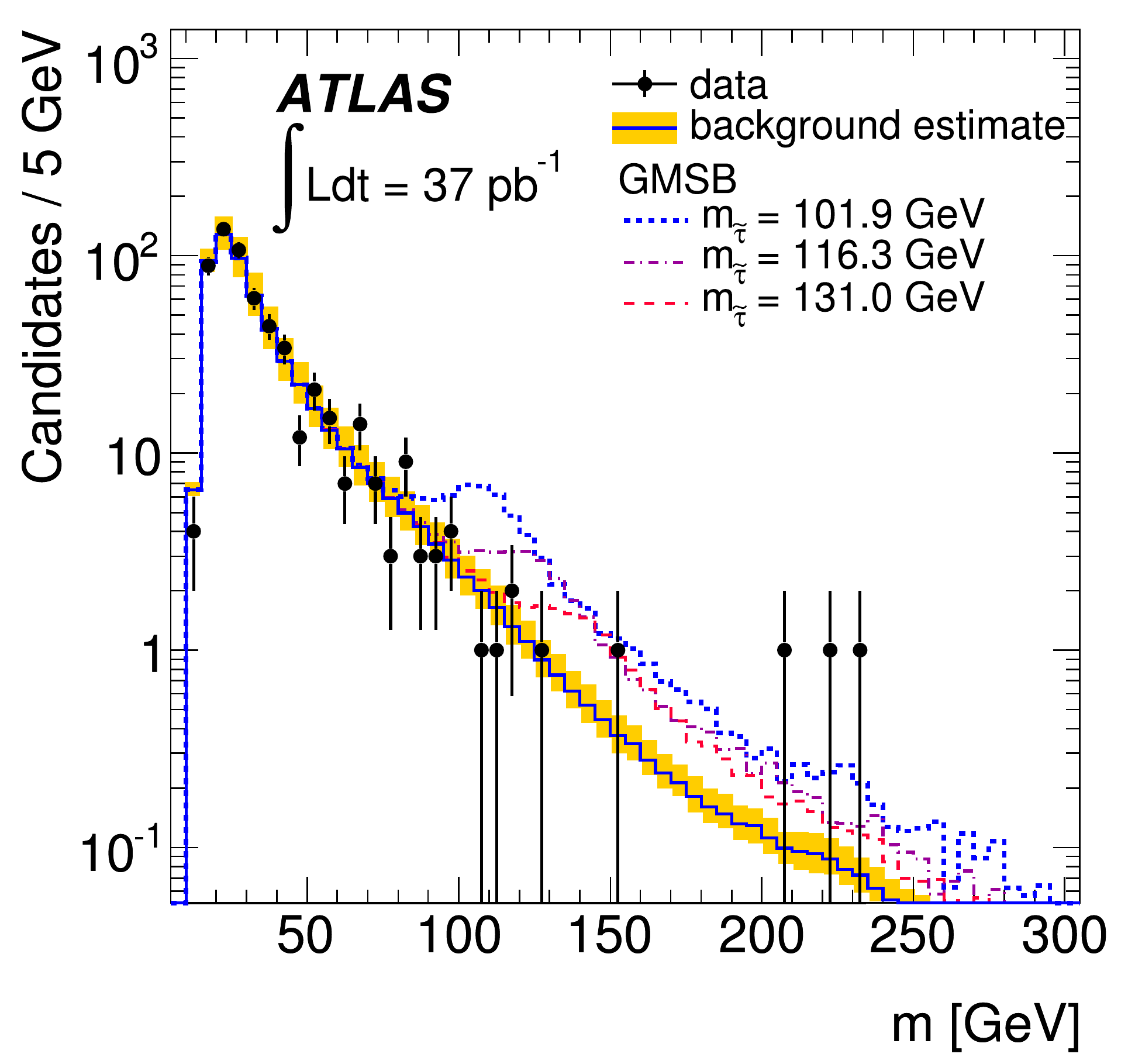}
  \includegraphics[width=0.9\columnwidth]{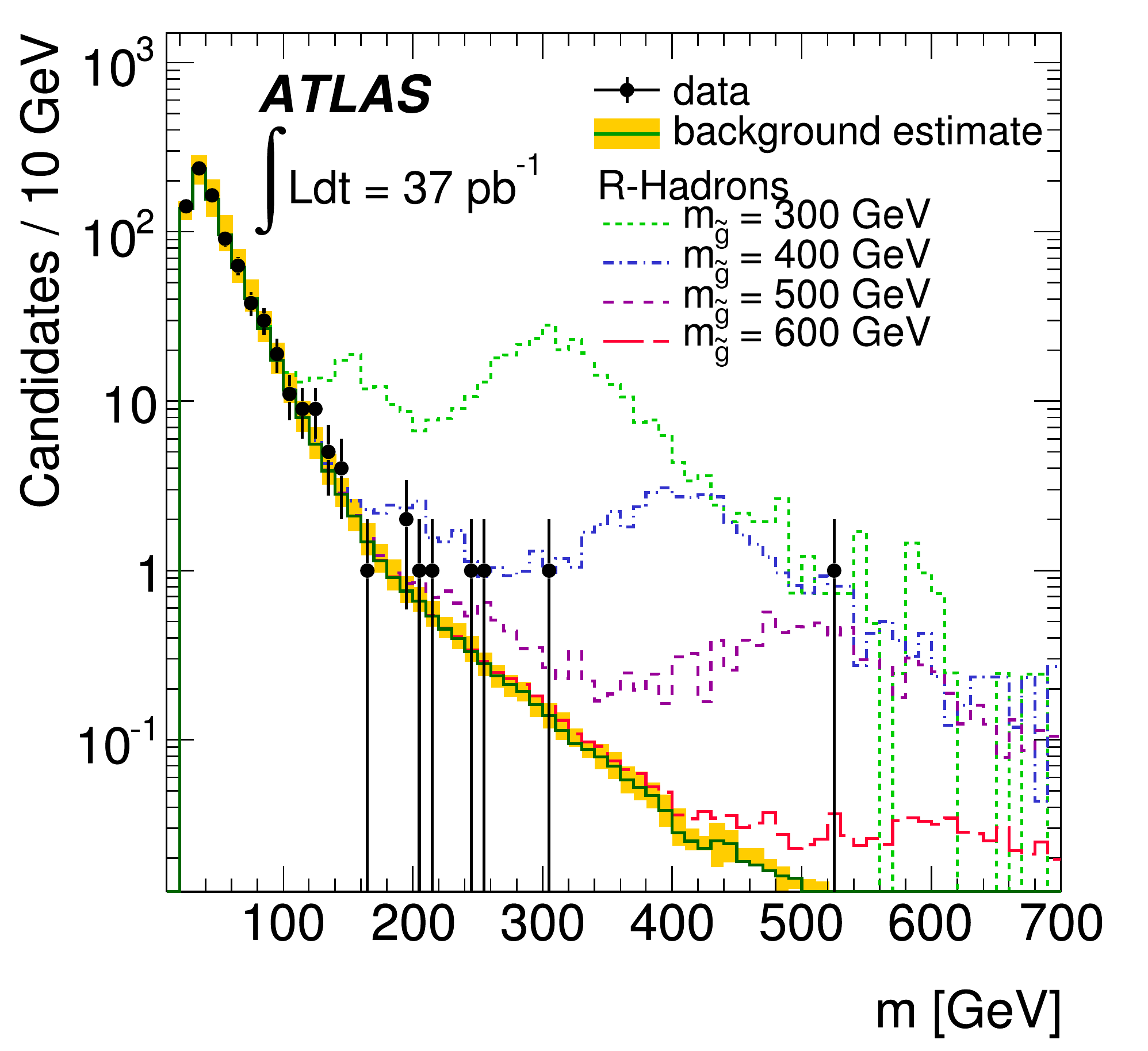}
  \caption{Candidate estimated mass distribution for data, expected background including systematic uncertainty, with simulated signals added, in the slepton (upper) and {\it{R}}-hadron (lower) 
searches~\cite{bib:llp2}.}
 \label{fig:signalOverEstimator}
\end{figure}
A complimentary approach~\cite{bib:llp2} relies on a signal being present in the muon spectrometer (MS). In this case a muon trigger allows for selection
of a relevant data sample and selections that taregt the lepton-like and hadron-like LLPs are approached separately.
For the lepton-like LLPs, two candidates with $p_{T}>$40~GeV are required, where each respective candidate must have a combined ID+MS
reconstructed track. For the hadron-like selection, only one LLP candidate is required with $p_T>$60~GeV is demanded per event, and where 
combined tracks are not evidenced, MS standalone tracks are also considered. This provides sensitivity to the $R$-hadrons that are neutral 
in the ID while attaining a charge prior to reaching the MS. The momentum criteria that are applied are driven by the desire to be operating
in the plateau region of the MS trigger. It should be noted that candidates with reconstructed transverse momenta greater than 1~TeV are rejected.
The background estimation is performed in a similar way as described in Section~\ref{subsec:muon_agnostic} 
and can be seen demonstrated for the lepton- and hadron-like selections respectively in Figure~\ref{fig:signalOverEstimator}. 
The $\beta$ distribution and $|\eta|$ range is different depending on the subsystem being considered. To minimize this effect, the background estimation
is performed in $\eta$ regions such that the resolution on $\beta$ within each region is approximately the same. 

For each candidate a single measurement of $\beta$ is calculated, using information from the precision monitored drift tube chambers, and the fast resistive plate
chambers in the MS. Where available, tile calorimeter information is used to supplement the muon detector measurements for improved resolution.
When combined with the track's momentum measurement a mass estimate is calculated for each candidate.

\begin{figure}[tbp]
\includegraphics[width=0.5\textwidth]{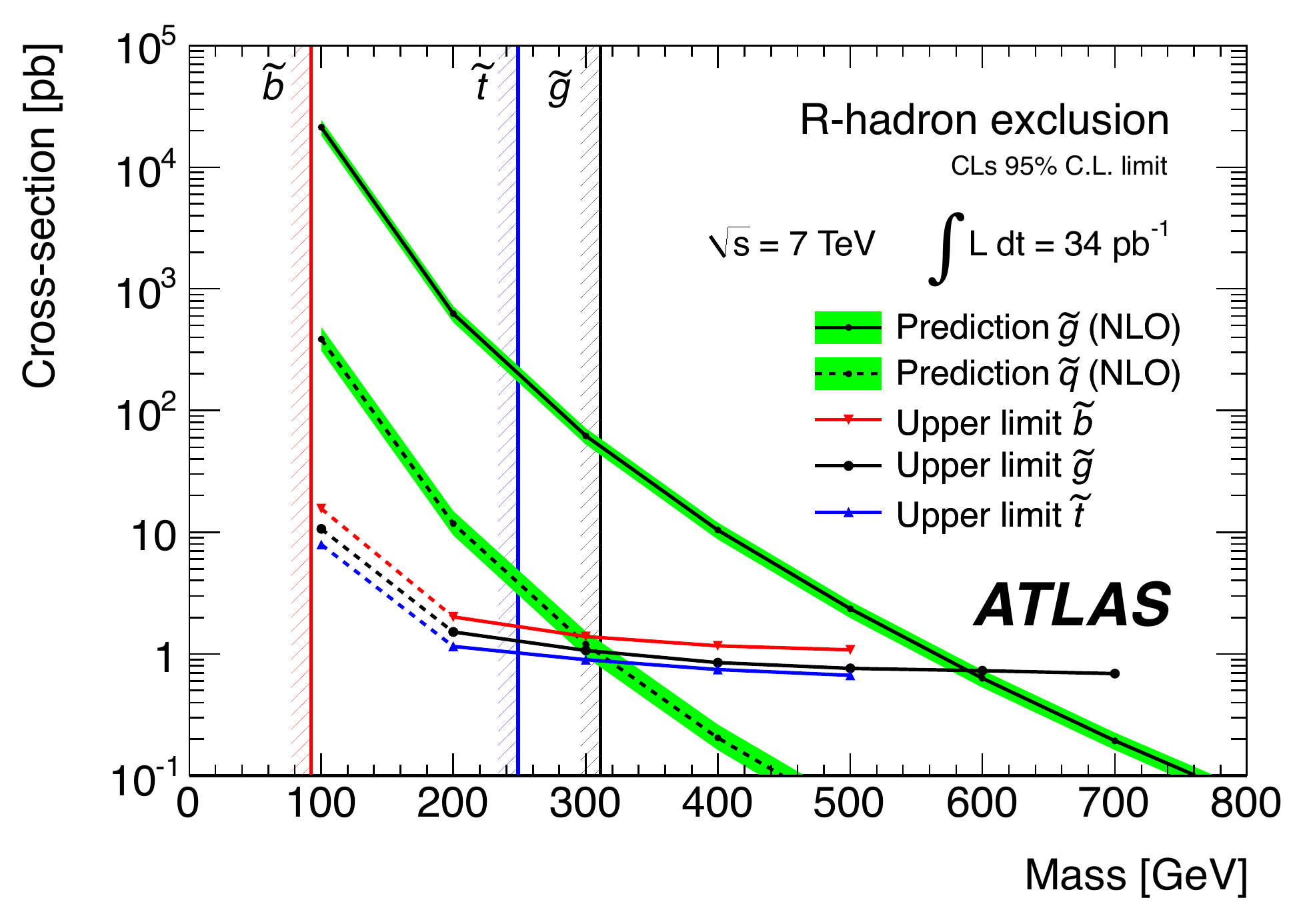}
\caption{Measured 95$\%$ C.L. cross section upper limits for $R$-hadrons~\cite{bib:llp1}, using the analysis as outlined in Section~\ref{subsec:muon_agnostic}.
\label{fig:llp1}}
\end{figure}
\begin{figure}[tbp]
\includegraphics[width=0.5\textwidth]{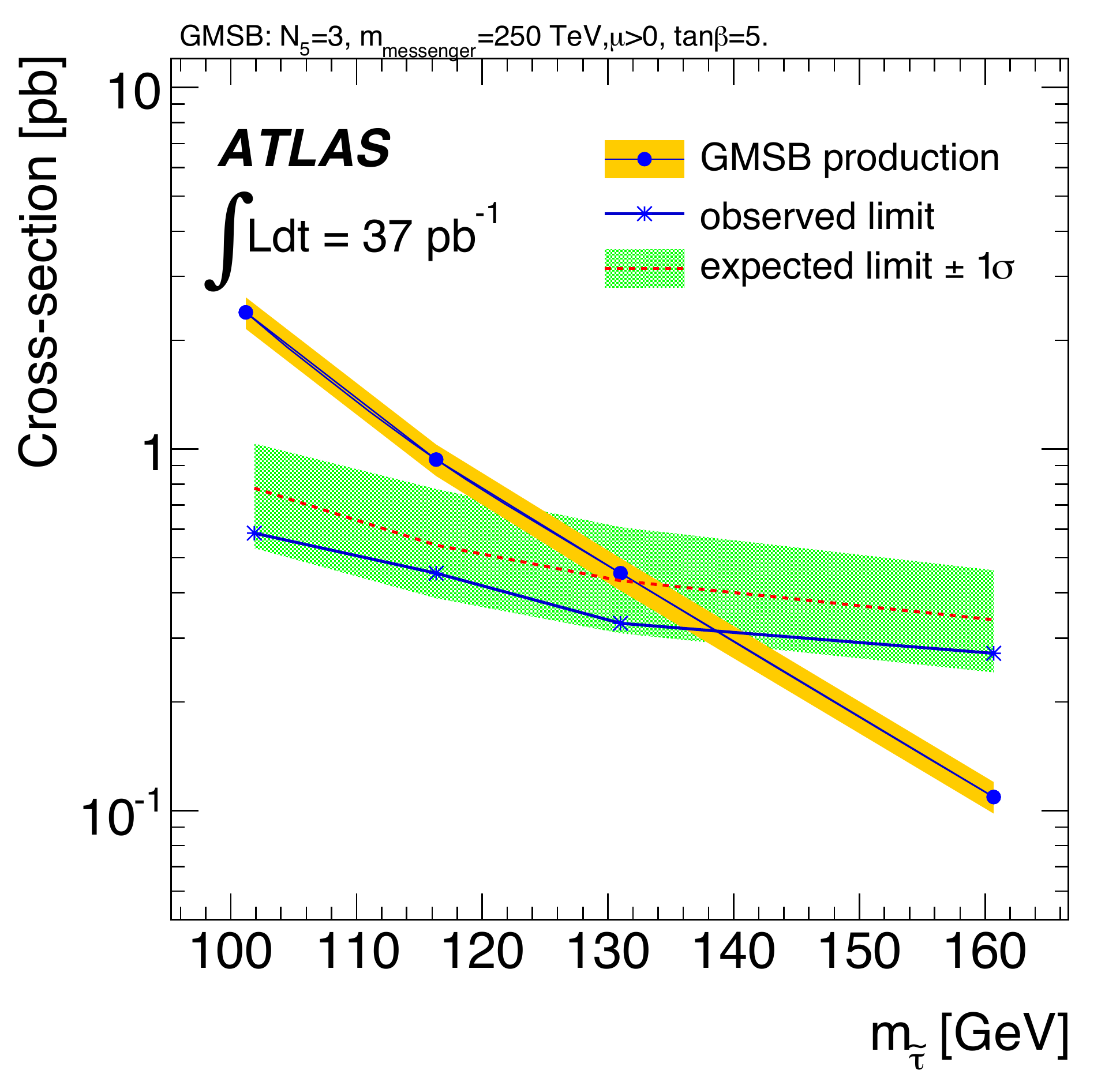}
\caption{Measured 95$\%$ C.L. cross section upper limits for long-lived sleptons~\cite{bib:llp2} using the analysis outlined in Section~\ref{subsec:muon}.
\label{fig:llp2}}
\end{figure}

\section{\label{sec:results}Results}

For each search strategy the observations agree well with low yields predicted due to instrumental effects in a background-only hypothesis.
Cross section upper limits, at the 95$\%$ C.L., are calculated for the production of each signal scenario. 
Figure~\ref{fig:llp1} illustrates constraints on the cross section in $R$-Hadron scenarios~\cite{bib:farrar}  
along with theoretical predictions for production of $\tilde{g}$, $\tilde{t}$ and $\tilde{b}$.
Constraints on a model of gauge mediated supersymmetry breaking featuring a stable slepton are shown in Figure~\ref{fig:llp2}. 
The intersections of the curves in each plot show the constraints with $m_{\tilde{g}}>$586~GeV, $m_{\tilde{t}}>$309~GeV, $m_{\tilde{b}}>$284~GeV, 
and $m_{\tilde{\tau}}>$136~GeV.

%%\bigskip % extra skip inserted

\end{document}